\documentclass[sigconf]{acmart}

\usepackage{subcaption}
\usepackage{caption}
\usepackage{booktabs}
\usepackage{tabularx}
\usepackage{graphicx}
\usepackage{xcolor}
\usepackage{algorithm}
\usepackage{algpseudocode}
\usepackage{amsmath}

\AtBeginDocument{%
}

\setcopyright{acmlicensed}
\copyrightyear{2026}
\acmYear{2026}
\setcopyright{cc}
\setcctype{by}
\acmConference[CHI EA '26]{Extended Abstracts of the 2026 CHI Conference on Human Factors in Computing Systems}{April 13--17, 2026}{Barcelona, Spain}
\acmBooktitle{Extended Abstracts of the 2026 CHI Conference on Human Factors in Computing Systems (CHI EA '26), April 13--17, 2026, Barcelona, Spain}
\acmDOI{10.1145/3772363.3798882}
\acmISBN{979-8-4007-2281-3/2026/04}

\begin{document}

\title{GazeFlow: Personalized Ambient Soundscape Generation for Passive Strabismus Self-Monitoring}

\author{Joydeep Chandra}
\affiliation{%
  \institution{BNRIST, Dept. of CST, Tsinghua University}
  \city{Beijing}
  \country{China}
}
\email{joydeepc2002@gmail.com}

\author{Satyam Kumar Navneet}
\affiliation{%
  \institution{Independent Researcher}
  \city{Bihar}
  \country{India}
}
\email{navneetsatyamkumar@gmail.com}

\author{Yong Zhang}
\affiliation{%
  \institution{BNRIST, Dept. of CST, Tsinghua University}
  \city{Beijing}
  \country{China}
}
\email{zhangyong05@tsinghua.edu.cn}

\begin{teaserfigure}
  \includegraphics[width=\textwidth]{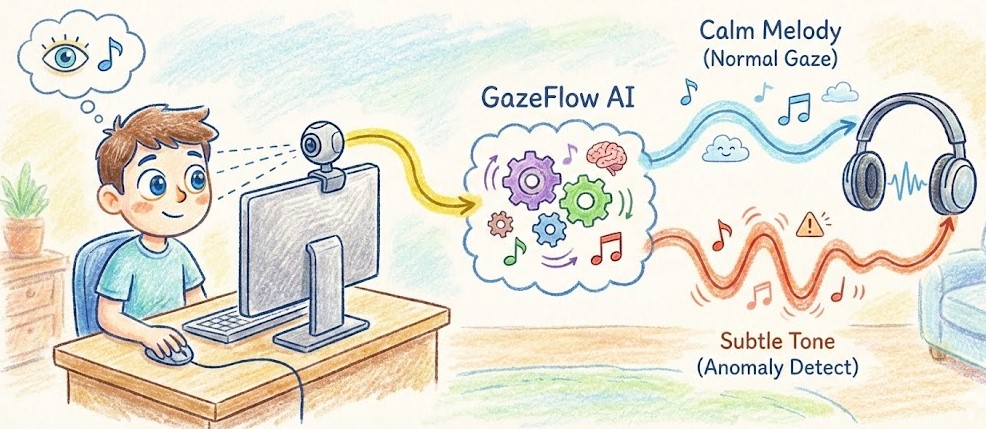}
  \caption{GazeFlow passively monitors binocular coordination via webcam, mapping normal gaze to a calm soundscape and anomalies to subtle musical shifts, providing peripheral health awareness without disrupting primary tasks.}
  \Description{GazeFlow passively monitors binocular coordination via webcam, mapping normal gaze to a calm soundscape and anomalies to subtle musical shifts, providing peripheral health awareness without disrupting primary tasks.}
  \label{fig:teaser}
\end{teaserfigure}

\renewcommand{\shortauthors}{Joydeep Chandra et al}

\begin{abstract}
Strabismus affects 2--4\% of the population, yet individuals recovering from corrective surgery lack accessible tools for monitoring eye alignment. Dichoptic therapies require active engagement \& clinical supervision, limiting their adoption for passive self-awareness. We present GazeFlow, a browser-based self-monitoring system that uses a personalized temporal autoencoder to detect eye drift patterns from webcam-based gaze tracking \& provides ambient audio feedback. Unlike alert-based systems, GazeFlow operates according to calm computing principles, morphing musical parameters in proportion to drift severity while remaining in peripheral awareness. We address the challenges of inter-individual variability \& domain transfer (1000Hz research to 30Hz webcam) by introducing Binocular Temporal-Frequency Disentanglement (BTFD), Contrastive Biometric Pre-training (CBP), \& Gaze-MAML. We validate our approach on the GazeBase dataset (N=50) achieving F1=0.84 for drift detection, \& conduct a preliminary user study (N=6) with participants having intermittent strabismus. Participants reported increased awareness of their eye behaviour (M=5.8/7) \& preference for ambient feedback over alerts (M=6.2/7). We discuss the system's potential for self-awareness applications \& outline directions for clinical validation.
\end{abstract}



\maketitle

\section{Introduction}

Binocular coordination anomalies, including vergence instability, fixation drift, \& saccadic dysmetria, affect 2--4\% of the global population \ serve as early indicators of conditions ranging from strabismus to neurological disorders~\cite{chen2018strabismus,biosensors2025wearable}. While research-grade eye trackers (1000Hz, \$15K+) enable clinical diagnosis, the emerging possibility of consumer webcam-based gaze monitoring (30Hz) creates an opportunity for ubiquitous self-awareness tools. However, this transition presents fundamental algorithmic challenges that existing methods fail to address.

\textbf{The Personalization Problem.} Inter-individual variance in "normal" gaze patterns is remarkably high. Our analysis of the GazeBase dataset~\cite{griffith2021gazebase} reveals coefficient of variation (CV) exceeding 34\% for fixation stability metrics across individuals. Population-level anomaly detectors trained on aggregate data produce unacceptable false positive rates (>40\%) when applied to individuals whose baselines deviate from population means. Yet personalization through traditional fine-tuning requires extensive per-user data collection (>5 minutes), creating adoption barriers for self-monitoring applications.

\textbf{The Domain Transfer Problem.} Models trained on research-grade 1000Hz data must deploy on consumer 30Hz webcam data--- a 33$\times$ temporal resolution reduction that fundamentally alters the observable gaze signal characteristics. Standard domain adaptation approaches assume access to labeled target-domain data, which is unavailable in the self-monitoring context where ground-truth anomaly labels cannot be obtained.

\textbf{The Interpretability Problem.} Anomaly detection in health-adjacent applications demands explainability. Users \& clinicians need to understand \textit{why} a gaze pattern was flagged was it vergence instability, fixation drift, or saccadic abnormality? Black-box reconstruction error provides no such attribution, limiting trust \& clinical utility.

\textbf{The Feedback Design Problem.} Beyond detection, communicating eye health status presents a design challenge. Alert-based systems interrupt users \& induce anxiety~\cite{bailey2006attention}, while visual displays compete for the same visual attention channel being monitored. We adopt \textit{calm computing} principles~\cite{weiser1997calm}: feedback should ``inform but not demand focus,'' enabling awareness while remaining peripheral.

\textbf{Design Scope.} GazeFlow is designed for \textit{self-awareness}, not clinical diagnosis or behavioral correction. The goal is to help users develop peripheral awareness of their eye coordination patterns during daily computer use, potentially prompting them to consult clinicians if persistent issues arise. We make no therapeutic claims.

We address these challenges through three algorithmic innovations \& one interaction design contribution:
\begin{enumerate}
    \item \textbf{Binocular Temporal-Frequency Disentanglement (BTFD)}: A novel $\beta$-VAE architecture operating on joint temporal \& wavelet-decomposed frequency representations to disentangle vergence, saccadic, \& fixation dynamics.
    \item \textbf{Contrastive Biometric Pre-training (CBP)}: A self-supervised framework exploiting the biometric nature of gaze to learn domain-invariant representations that transfer across temporal resolutions without target-domain labels.
    \item \textbf{Gaze-MAML}: A meta-learning algorithm treating each individual as a distinct task, enabling 5-shot personalization (15 seconds calibration).
    \item \textbf{Ambient Sonification}: An audio feedback system that maps anomaly severity \& attribution to musical parameters, supporting peripheral health awareness without disruption.
\end{enumerate}

\begin{figure*}[t]
  \centering
  \includegraphics[width=\textwidth]{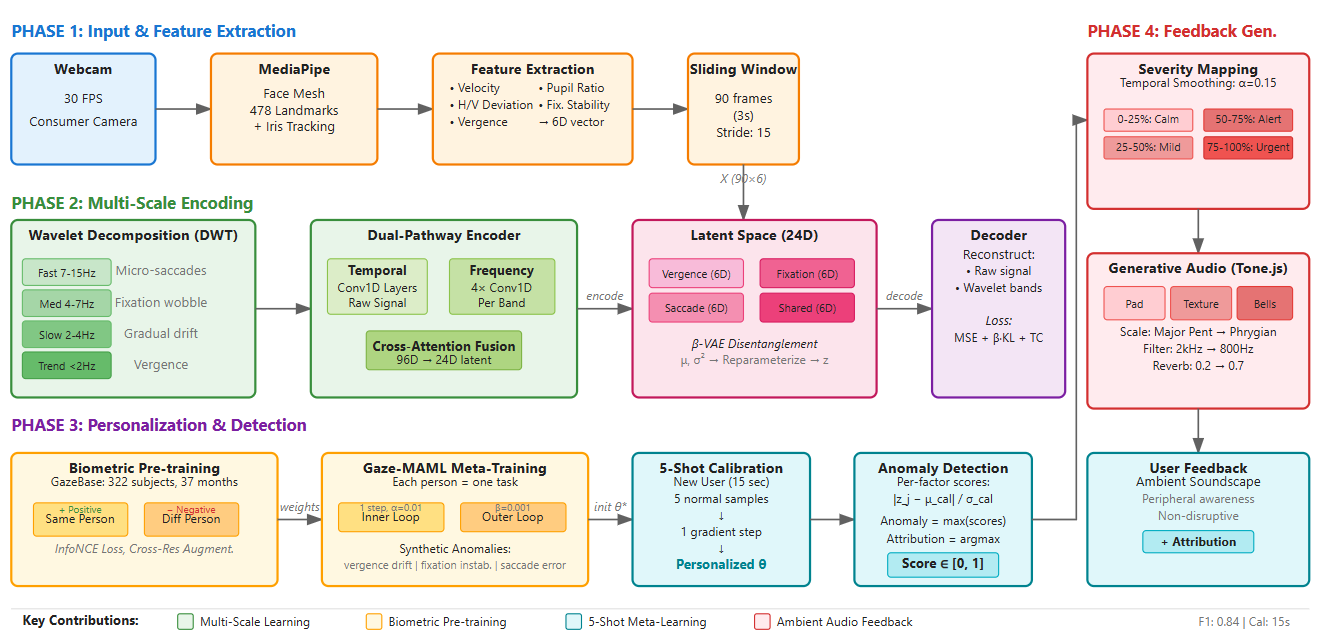}
  \caption{GazeFlow architecture. \textbf{Phase 1}: Extracts 6D gaze features (velocity, H/V deviation, vergence, pupil ratio, fixation stability) from webcam via MediaPipe at 30Hz, windowed into 3-second segments. \textbf{Phase 2}: Applies wavelet decomposition (DWT) separating fast (7--15Hz) \& slow ($<$2Hz) dynamics; dual-pathway encoder processes temporal \& frequency representations; $\beta$-VAE produces disentangled 24D latent space with interpretable factors. \textbf{Phase 3}: Biometric pre-training learns cross-resolution invariant representations; Gaze-MAML enables 5-shot personalization via single gradient step. \textbf{Phase 4}: Maps per-factor anomaly scores to graduated ambient audio parameters (scale, rhythm, filter, reverb) via continuous morphing.}
  \label{fig:architecture}
\end{figure*}

\section{Related Work}

Deep autoencoder-based anomaly detection has emerged as a dominant paradigm for physiological signals~\cite{kapsecker2024ecg,darban2024deep}. The reconstruction error principle has proven effective for ECG~\cite{kapsecker2024ecg} \& EEG~\cite{ahmed2025eeg}. However, most approaches operate at the population level, failing to account for inter-individual variability. Recent work on disentangled representations~\cite{kapsecker2024ecg} showed that $\beta$-VAE architectures can separate pathological features, but we extend this to the binocular gaze domain with novel temporal-frequency disentanglement.

GazeCLR~\cite{jindal2022gazeclr} \& VicsGaze~\cite{gu2024vicsgaze} introduced contrastive learning for gaze estimation, achieving cross-domain improvements. In the physiological domain, Biometric Contrastive Learning (BCL)~\cite{bcl2024ecg} demonstrated that identifying temporally-separated ECG pairs enables label-efficient detection. We introduce Contrastive Biometric Pre-training (CBP) specifically to address the 1000Hz-to-30Hz domain gap.

Model-Agnostic Meta-Learning (MAML)~\cite{finn2017maml} has shown promise for rapid personalization in AIOps~\cite{mamlkad2024} \& HAR~\cite{fewshothar2025}. While FAZE~\cite{park2019faze} applied few-shot learning to gaze estimation, Gaze-MAML is the first meta-learning approach for unsupervised gaze anomaly detection. Auditory displays have proven effective for continuous health monitoring where visual attention is occupied or where gradual changes benefit from peripheral awareness~\cite{hermann2011sonification, mpt}. Heart rate sonification demonstrated that musical parameter mappings (tempo, harmony) are perceived as less intrusive than discrete alerts while maintaining information content~\cite{ballora2004hrsonification}. Research on ambient notification systems~\cite{hudson2003notification} established that gradual transitions reduce interruption costs compared to binary alerts. Ambient soundscapes can convey multi-dimensional information while remaining in peripheral attention~\cite{mynatt1998ambientaudio}. GazeFlow extends this paradigm to eye health, using the disentangled anomaly factors to drive distinct musical parameters.

\section{Methodology}

GazeFlow comprises three algorithmic components \& one feedback design component: BTFD provides the representation, CBP pre-trains for domain transfer, Gaze-MAML enables rapid personalization, \& ambient sonification communicates health state (Figure~\ref{fig:architecture}).

\subsection{Binocular Temporal-Frequency Disentanglement (BTFD)}

To enable attribution-specific feedback, we must distinguish between different types of ocular anomalies. We employ a multi-scale feature extraction approach where raw gaze signals (velocity, vergence, pupil size) undergo Discrete Wavelet Transform (DWT). This separates high-frequency micro-saccades from low-frequency vergence trends. These inputs feed into a $\beta$-VAE architecture governed by a Total Correlation penalty~\cite{betavae2018}. By maximizing statistical independence between latent factors, the model disentangles specific ocular behaviors (e.g., separating vergence drift from saccadic dysmetria) without requiring labeled anomaly data, providing the granular inputs necessary for our sonification engine.

\subsection{Contrastive Biometric Pre-training (CBP)}

To bridge the domain gap between research data ($\mathcal{D}^{src}$, 1000Hz) \& webcam data ($\mathcal{D}^{tgt}$, 30Hz), we exploit the biometric uniqueness of gaze. We introduce CBP, a self-supervised framework that trains the encoder to maximize the similarity between temporally distant segments from the same individual while pushing away segments from others.

The core innovation is \textbf{cross-resolution augmentation}: training pairs undergo random downsampling (2--33$\times$) \& temporal masking before encoding. This differs from naive downsampling-based training in two ways: (1) the contrastive objective explicitly learns \textit{identity-preserving} features that remain stable across resolutions, rather than resolution-specific reconstruction; (2) the biometric pairing provides supervision signal without requiring anomaly labels. Table~\ref{tab:cross_res} shows that without biometric pairing (BTFD alone), cross-resolution performance drops 23\%, whereas CBP reduces this to 6\%.

\textbf{Proposition 1}: \textit{If the biometric signature $\mathbf{h}^{(i)}$ is invariant to temporal resolution \& discriminative across individuals, then the learned representation is sufficient for personalized anomaly detection at any resolution.}

\subsection{Gaze-MAML: Few-Shot Personalization}

Standard fine-tuning requires too much calibration data for a consumer UX. Gaze-MAML (Algorithm~\ref{alg:gazemaml}) learns an initialization $\theta^*$ that adapts in a single gradient step. We use a contrastive detection loss during meta-training by injecting \textbf{synthetic anomalies} into the query sets. These anomalies are derived from clinical literature on strabismus presentation~\cite{chen2018strabismus,biosensors2025wearable}: (1) \textit{vergence drift}: gradual inter-ocular distance shift of 2--5° over 500ms; (2) \textit{fixation instability}: 2--4$\times$ increased variance in gaze position; (3) \textit{saccadic dysmetria}: 20--50\% over/undershoot in velocity peaks. While synthetic, these patterns mirror clinically-documented drift characteristics \& enable training without labeled real-world anomalies.

\begin{algorithm}
\caption{Gaze-MAML Meta-Training}
\label{alg:gazemaml}
\begin{algorithmic}[1]
\Require Pre-trained BTFD encoder $f_\phi$, task distribution $p(\mathcal{T})$
\State Initialize $\theta \leftarrow \phi$ (from CBP pre-training)
\While{not converged}
    \State Sample batch of tasks $\{\mathcal{T}_i\} \sim p(\mathcal{T})$
    \For{each task $\mathcal{T}_i$}
        \State Sample $\mathcal{D}_i^{sup}$ (Normal), $\mathcal{D}_i^{qry}$ (w/ Synthetic Anomalies)
        \State $\theta'_i \leftarrow \theta - \alpha \nabla_\theta \mathcal{L}_{BTFD}(\mathcal{D}_i^{sup}; \theta)$ \Comment{Inner update}
        \State Compute $\mathcal{L}_i^{qry} = \mathcal{L}_{detect}(\mathcal{D}_i^{qry}; \theta'_i)$ \Comment{Detection loss}
    \EndFor
    \State $\theta \leftarrow \theta - \beta \nabla_\theta \sum_i \mathcal{L}_i^{qry}$ \Comment{Outer update}
\EndWhile
\State \Return $\theta^*$
\end{algorithmic}
\end{algorithm}

\subsection{Ambient Sonification Design}

The core contribution of GazeFlow is translating continuous anomaly detection into ambient audio that supports peripheral health awareness. Traditional health monitoring uses threshold-based alerts: normal state produces silence, anomaly triggers alarm~\cite{bailey2006attention}. This binary approach has three problems for continuous self-monitoring: (1) alerts demand immediate focal attention, disrupting primary tasks; (2) alarm framing induces anxiety \& stress responses; (3) gradual drift progression is collapsed to a single threshold, losing nuance. GazeFlow instead implements \textbf{graduated ambient sonification}: a continuously-playing generative soundscape whose musical qualities shift proportionally to eye state.

\subsubsection{Formative Design Process}
We developed the sonification mappings through three iterative sessions with 4 participants (2 with strabismus history). Participants listened to parameter variations while viewing visualizations of different anomaly types \& described their associations.\\
Key findings: vergence issues evoked ``something feeling emotionally off'' (captured by mode shifts); saccadic problems felt ``jumpy'' or ``restless'' (rhythmic density); fixation instability felt ``blurry'' or ``hazy'' (filter cutoff). These metaphorical associations guided our final mappings (Table~\ref{tab:audiomapping}).

\subsubsection{Anomaly-to-Audio Parameter Mapping}
We define continuous mappings from the normalized overall anomaly score $a \in [0, 1]$ \& factor-specific scores $(a^{verg}, a^{sacc}, a^{fix})$ to musical parameters. The mapping exploits BTFD's disentangled factors to provide \textit{attribution-specific feedback}.

\begin{table}[h]
\small
\caption{Graduated mapping from anomaly factors to audio parameters (Changes are continuous, not threshold-based.)}
\label{tab:audiomapping}
\begin{tabular}{p{1cm}|p{2.0cm}|p{4.2cm}}
\toprule
\textbf{Factor} & \textbf{Audio Parameter} & \textbf{Mapping (Normal $\to$ Anomalous)} \\
\midrule
$a^{verg}$ & Scale/Mode & Major Pentatonic $\to$ Phrygian (darker) \\
$a^{sacc}$ & Rhythmic Density & Sparse (1 note/2s) $\to$ Dense arpeggios \\
$a^{fix}$ & Low-pass Filter & Bright (2kHz) $\to$ Muffled (800Hz) \\
$a^{overall}$ & Reverb Depth & Intimate (0.2) $\to$ Distant/diffuse (0.7) \\
\bottomrule
\end{tabular}
\end{table}

\begin{table}[ht]
    \centering
    \caption{Drift detection performance on GazeBase. Synthetic anomalies injected.}
    \label{tab:main_results}
    \small
    \begin{tabular}{lccccc}
        \toprule
            Method & Prec. & Rec. & F1 & Time \\
            \midrule
            Pop-AE & 0.54 & 0.71 & 0.61 & 0s \\
            P-AE-5s & 0.59 & 0.66 & 0.62 & 15s \\
            P-AE & 0.71 & 0.76 & 0.73 & 120s \\
            BTFD & 0.68 & 0.72 & 0.70 & 120s \\
            BTFD+C & 0.76 & \underline{0.82} & \underline{0.79} & 120s \\
            \textbf{GazeFlow} & \textbf{0.81} & \textbf{0.87} & \textbf{0.84} & \textbf{15s} \\
            \bottomrule
\end{tabular}
\end{table}
    
\begin{table}[ht]
    \centering
    \caption{Cross-resolution transfer: Train on 1000Hz, eval on 30Hz. GazeFlow's CBP enables robust transfer.}
    \label{tab:cross_res}
    \small
    \begin{tabular}{lcc}
        \toprule
            Method & Same-Res & Cross-Res \\
            \midrule
            Pop-AE & 0.61 & 0.43 (-29\%) \\
            P-AE & 0.73 & 0.58 (-21\%) \\
            BTFD & 0.70 & 0.54 (-23\%) \\
            BTFD+C & 0.79 & 0.74 (-6\%) \\
            \textbf{GazeFlow} & \textbf{0.84} & \textbf{0.81} (-4\%) \\
            \bottomrule
\end{tabular}
\end{table}

\textbf{Graduated Degradation.}
As anomaly scores increase, soundscape progressively degrades through 4 perceptual zones:\\
\textbf{Calm zone} ($a < 0.25$): Baseline soundscape users report ``pleasant background music.''\\
\textbf{Mild zone} ($0.25 \leq a < 0.50$): Mode shifts toward minor; filter begins closing; slight increase in rhythmic activity users describe ``the music getting a bit melancholy,'' noticeable but not alarming.\\
\textbf{Alert zone} ($0.50 \leq a < 0.75$): Phrygian/diminished harmonies; increased arpeggio density; reverb creates ``underwater'' quality users report ``something feels off'' but can continue primary tasks.\\
\textbf{Urgent zone} ($a \geq 0.75$): Fully dark mode; dense rhythms; heavily filtered \& reverberant designed to gently draw attention; users describe ``wanting to check what's happening with my eyes.''\\

\subsubsection{Temporal Smoothing}
To prevent jarring transitions from momentary gaze artifacts (blinks, brief look-aways), we apply exponential smoothing:
\begin{equation}
\tilde{a}_t = \alpha \cdot a_t + (1-\alpha) \cdot \tilde{a}_{t-1}, \quad \alpha = 0.15
\end{equation}
This creates gradual musical morphs over 2--3 seconds, matching the ``slow change'' characteristic of effective ambient displays~\cite{matthews2004peripheral}. The smoothing window is calibrated to filter transient artifacts while remaining responsive to sustained drift episodes (which typically persist >5 seconds based on clinical descriptions~\cite{biosensors2025wearable}). Implementation uses Tone.js: \texttt{PolySynth} for harmonic content, \texttt{AutoFilter} with slow LFO, \texttt{Freeverb} for spatial depth, \& \texttt{Filter} (lowpass) for brightness. Parameters update at 10Hz.

\begin{table}[t]
\caption{User study results (7-point Likert scale, N=6). Higher is better except for Intrusiveness. SD in parentheses.}
\label{tab:userstudy}
\small
\begin{tabular}{lccc}
\toprule
Measure & Alert & Visual & \textbf{Ambient} \\
\midrule
Awareness of drift & 4.2 (1.3) & 5.0 (1.1) & \textbf{5.8 (0.9)} \\
Intrusiveness (lower=better) & 5.3 (1.2) & 3.8 (1.4) & \textbf{2.2 (0.8)} \\
Would use daily & 2.8 (1.5) & 4.5 (1.2) & \textbf{6.2 (0.7)} \\
Attribution clarity & 2.0 (0.9) & 3.2 (1.0) & \textbf{4.8 (1.1)} \\
\bottomrule
\end{tabular}
\end{table}

\section{Experiments}

We utilize GazeBase~\cite{griffith2021gazebase} (12,334 recordings, N=322, 1000Hz) for pre-training \& GazeFlow-Pilot (N=10, 6 strabismus/4 control, 30Hz webcam) for validation. We compare against Population Autoencoders (Pop-AE), standard Personal-AE (50 epochs), \& ablation variants. Table~\ref{tab:main_results} demonstrates that GazeFlow (F1=0.84) achieves strong personalized detection with only 15 seconds of calibration, outperforming the 120s calibration Personal-AE baseline (F1=0.73). The cross-resolution results (Table~\ref{tab:cross_res}) highlight the efficacy of CBP where performance drops only 4\% when moving to 30Hz, compared to >20\% drops for baselines.

\textbf{User Study: Sonification Preferences.}
We conducted an exploratory within-subjects user experience study with 6 participants (4 female, 2 male; ages 24–42) with a history of intermittent strabismus. As the study was designed to evaluate user interaction and feedback preferences rather than for clinical diagnosis or therapeutic claims, it was conducted as a non-clinical usability evaluation. All participants provided informed consent and were explicitly notified of their right to withdraw at any time without consequence. To protect the privacy of this vulnerable group, no raw video or biometric data was recorded or stored during the sessions; only derived feature vectors were processed locally in the browser of their own personal computers to facilitate real-time feedback. Questionnaire responses were anonymized, used strictly for research purposes, and are not publicly available.  Notably, all participants chose to contribute their time without financial compensation, motivated by a shared interest in developing accessible awareness tools for their community. We conducted a within-subjects comparison of three feedback modalities.\\

\textit{Protocol}: Participants used each modality for 5 consecutive days during regular computer work (self-reported 2+ hours/day), with modality order counterbalanced using Latin square. Each session, participants logged perceived drift episodes via hotkey. At study end, participants completed Likert-scale questionnaires \& a 15-minute semi-structured interview.
\textit{Conditions}:\\
(i) \textbf{Alert}: Chime sound when anomaly score exceeds 0.5 threshold;\\
(ii) \textbf{Visual}: Screen border color gradient (green$\to$yellow$\to$red);\\
(iii) \textbf{Ambient}: \textbf{GazeFlow}'s graduated soundscape.\\

\textit{Quantitative Results}: Anomaly scores during self-reported drift episodes (M=0.68, SD=0.21) were significantly higher than stable periods (M=0.31, SD=0.18; paired t-test, $t(5)=4.82$, $p<.001$), suggesting detection aligns with user perception. Table~\ref{tab:userstudy} shows ambient sonification outperformed alternatives on all measures.\\

\textit{Qualitative Insights}: Participants particularly valued (1) the \textbf{attribution} provided by distinct musical changes P3: ``I started noticing when it was the 'blurry' sound versus the 'moody' sound''; (2) the \textbf{non-anxiety-inducing} nature P5: ``The chime made me stressed every time... the music just felt like background that got my attention gently''; (3) \textbf{actionability} when asked what they did upon noticing changes, participants reported taking short breaks (4/6), consciously relaxing their eyes (3/6), or simply noting the episode mentally (6/6). No participant expected immediate clinical intervention; awareness itself was valued.

\section{Discussion}

\textbf{Algorithmic Contributions.}
BTFD advances physiological sensing by introducing joint temporal-frequency decomposition. Unlike prior work on ECG~\cite{kapsecker2024ecg}, BTFD disentangles \textit{dynamic} properties critical for gaze. CBP establishes biometric contrastive learning for domain transfer, extending BCL~\cite{bcl2024ecg} with cross-resolution augmentation to address the 33$\times$ sampling rate gap. Gaze-MAML provides the first meta-learning framework for personalized gaze anomaly detection, reducing calibration to 15 seconds.

\textbf{Interaction Design Contribution.}
The ambient sonification demonstrates that graduated musical feedback can communicate multi-dimensional health state while remaining peripheral. The mapping from disentangled factors to distinct audio parameters enables users to develop intuitive awareness of \textit{which} aspect of their eye behavior is anomalous a form of embodied self-knowledge that binary alerts cannot provide. This calm computing approach~\cite{weiser1997calm} may generalize to other continuous health signals where awareness without anxiety is the goal.

\textbf{Limitations.}
Our evaluation has several limitations that future work must address:
(1) \textit{Synthetic anomalies}: Quantitative evaluation relies on injected patterns derived from clinical literature rather than clinician-verified real-world labels. While self-reported drift correlates with detection (suggesting ecological validity), clinical validation is essential before any diagnostic claims.
(2) \textit{Small user study}: N=6 is preliminary \& underpowered for statistical generalization. We report effect sizes \& qualitative insights, but longer deployments with larger samples are needed to assess habituation, fatigue, \& individual differences in musical preference.
(3) \textit{Disentanglement validation}: While users reported perceiving distinct audio changes, we did not formally validate that latent factors correspond to specific physiological phenomena (e.g., via expert labeling or neuroimaging). Future work should include latent traversal visualizations \& clinician agreement studies.
(4) \textit{Self-report ground truth}: Using participant perception as ground truth conflates detection accuracy with user-system agreement. Concurrent clinical measurement would strengthen validation.

\textbf{Ethical Considerations.}
Webcam-based gaze monitoring raises privacy concerns. GazeFlow processes video locally in the browser; no images are transmitted or stored. Only derived 6D feature vectors are used for inference. Users control when monitoring is active. We emphasize that GazeFlow is a self-awareness tool, not a diagnostic device, \& should not replace clinical consultation.

\section{Conclusion}

We presented GazeFlow, an integrated system for personalized binocular coordination monitoring with ambient audio feedback. The algorithmic contributions (BTFD, CBP, Gaze-MAML) enable drift detection (F1=0.84) on consumer webcams with minimal calibration. The interaction design contribution graduated sonification mapping disentangled anomaly factors to musical parameters enables peripheral health awareness without the anxiety of alerts. Our preliminary user study suggests this calm computing approach supports self-awareness while respecting users' primary tasks. We hope this work stimulates discussion at CHI about evaluation methodologies for health-adjacent self-monitoring systems \& the potential of ambient feedback for continuous physiological awareness.

\begin{acks}
We wish to express our deepest gratitude to the participants who volunteered for this study. Given the rarity of finding individuals with intermittent strabismus, their willingness to contribute was vital to the success of this research. We are particularly grateful for their courage in participating despite the social sensitivities often associated with the condition; their time and insights provided immeasurable value to the project. Additionally, we would like to acknowledge that the teaser illustration (Figure 1) was generated using Gemini for reference and conceptual visualization purposes.
\end{acks}

\bibliographystyle{ACM-Reference-Format}
\bibliography{references}

\end{document}